\documentclass[9pt,conference]{IEEEtran}
\IEEEoverridecommandlockouts
\usepackage{cite}
\usepackage{tikz}
\usepackage{fancyhdr}
\usepackage{amsmath,amssymb,amsfonts}
\usepackage{algorithmic}
\usepackage{graphicx}
\usepackage{textcomp}
\usepackage{pifont}
\usepackage{xcolor}
\usepackage{wasysym}
\usepackage{flushend} 
\usepackage{multirow}
\usepackage[normalem]{ulem}
\useunder{\uline}{\ul}{}
\usepackage{acronym}
\usepackage[citecolor=black,
            colorlinks=true,
            linkcolor=black,
            urlcolor  = black,
            pdftitle={Speaker Embedding Informed Audiovisual Active Speaker Detection for Egocentric Recordings},
            pdfauthor={Jason Clarke, Yoshihiko Gotoh, and Stefan Goetze},
            pdfkeywords={Diarization, Audiovisual Active Speaker Detection, Video-based Face Recognition, Speaker Recognition},
            pdfsubject={Diarization, Audiovisual Active Speaker Detection, Video-based Face Recognition, Speaker Recognition}]{hyperref}
\usepackage[english]{babel}
\addto\extrasenglish{}
\addto\extrasenglish{}
\addto\extrasenglish{}
\addto\extrasenglish{}
\addto\extrasenglish{}
\addto\extrasenglish{}
\acrodef{AR}    {augmented reality}
\acrodef{ASD}   {audiovisual active speaker detection}
\acrodef{AVD}   {audiovisual diarization}
\acrodef{GRU}   {gated recurrent unit}
\acrodef{CP}    {conversation Participant}
\acrodef{DeiT}  {data-efficient image transformer}
\acrodef{FoV}   {field of view}
\acrodef{mAP}   {mean Average Precision}
\acrodef{MFCC}  {Mel frequency cepstral coefficient}
\acrodef{MLP}   {multilayer perceptron}
\acrodef{SCAN}   {speaker comparison auxiliary network} 
\acrodef{SNR}   {signal-to-noise-ratio}
\acrodef{SoTA}  {state-of-the-art}
\acrodef{V-TCN} {visual temporal convolutional network}

\def\BibTeX{{\rm B\kern-.05em{\sc i\kern-.025em b}\kern-.08em
    T\kern-.1667em\lower.7ex\hbox{E}\kern-.125emX}}

\newcommand\copyrighttext{%
\footnotesize \textcopyright 2025 IEEE. Personal use of this material is permitted.
Permission from IEEE must be obtained for all other uses, in any current or future
media, including reprinting/republishing this material for advertising or promotional
purposes, creating new collective works, for resale or redistribution to servers or
lists, or reuse of any copyrighted component of this work in other works.}

\newcommand\copyrightnotice{%
\begin{tikzpicture}[remember picture,overlay]
\node[anchor=south,yshift=10pt] at (current page.south)
{\fbox{\parbox{\dimexpr\textwidth-\fboxsep-\fboxrule\relax}{\copyrighttext}}};
\end{tikzpicture}%
}    
\begin{document}

\title{Speaker Embedding Informed Audiovisual Active Speaker Detection for Egocentric Recordings\\
\copyrightnotice

\thanks{This work was supported by the Centre for Doctoral Training in Speech and Language Technologies (SLT) and their Applications funded by UKRI [grant number EP/S023062/1]. This work was also funded in part by Meta.}
}


\author{
  \IEEEauthorblockN{
    Jason Clarke$^1$, 
    Yoshihiko Gotoh$^1$, and 
    Stefan Goetze$^{1,2}$
  }
  \IEEEauthorblockA{
    $^1$Speech and Hearing (SPandH) group, School of Computer Science,
    The University of Sheffield, Sheffield, United Kingdom
  }
  \IEEEauthorblockA{
    $^2$South Westphalia University of Applied Sciences, Iserlohn, Germany
  }
  \IEEEauthorblockA{
    \{jclarke8, y.gotoh, s.goetze\}@sheffield.ac.uk, goetze.stefan@fh-swf.de
  }
}

\maketitle

\begin{abstract}
\Ac{ASD} addresses the task of determining the speech activity of a candidate speaker given acoustic and visual data. Typically, systems model the temporal correspondence of audiovisual cues, such as the synchronisation between speech and lip movement. Recent work has explored extending this paradigm by additionally leveraging speaker embeddings extracted from candidate speaker reference speech. This paper proposes the \ac{SCAN} which uses speaker-specific information from both reference speech and the candidate audio signal to disambiguate challenging scenes when the visual signal is unresolvable. Furthermore, an improved method for enrolling face-speaker libraries is developed, which implements a self-supervised approach to video-based face recognition. Fitting with the recent proliferation of wearable devices, this work focuses on improving speaker-embedding-informed \ac{ASD} in the context of egocentric recordings, which can be characterised by acoustic noise and highly dynamic scenes. \ac{SCAN} is implemented with two well-established baselines, namely TalkNet and Light-ASD; yielding a relative improvement in mAP of $14.5\%$ and $10.3\%$ on the Ego4D benchmark, respectively.

\end{abstract}


\section{Introduction}
\Acf{ASD} revolves around determining the video-framewise speech activity of a candidate speaker. The task is typically formulated as a binary classification problem, where, given a mixed audio signal and sequence of temporally contiguous bounding boxes centered on the candidate speaker's face, a system identifies video frames where the candidate speaker is talking~\cite{talknet, 3DResNet, activespeakersincontext,asdtransformer,buffy,LeonAlcazar2021MAASMA}.

\begin{figure}[!ht]
  \centering
    \includegraphics[scale=0.05]{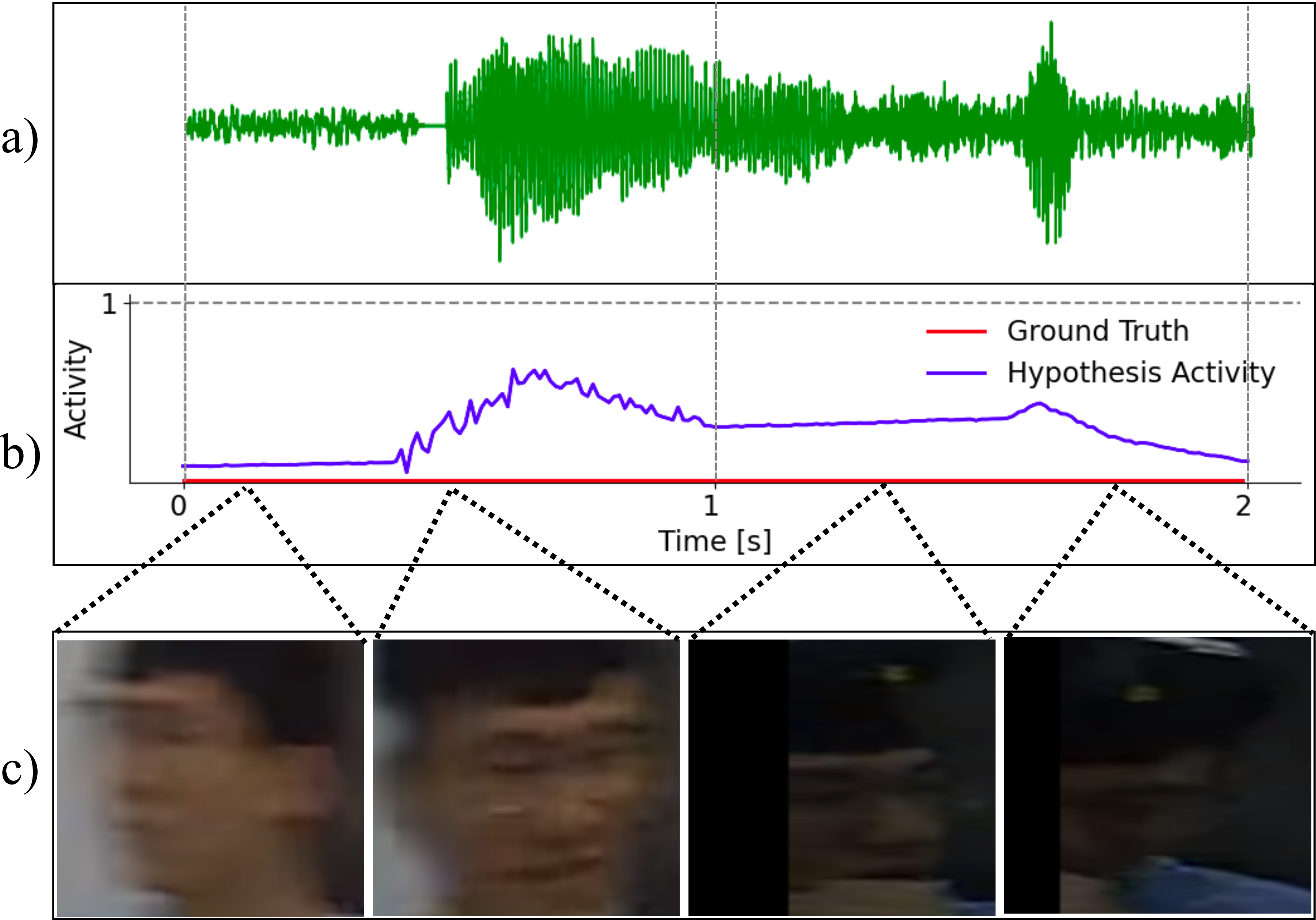}
  \caption{
  Example of typical false-positive \ac{ASD}: a) input audio signal; b) ground truth speaker activity of the candidate speaker (inactive throughout) and hypothesised speaker activity by a state-of-the-art speaker-embedding naive \ac{ASD} system~\cite{Liao_2023_CVPR}; c) selection of challenging video frames from a typical egocentric video track~\cite{Ego4D}.}
\label{fig:timeline}
\end{figure}

Previous \ac{ASD} research has mainly focused on improving performance for exocentric data (recorded from the third person perspective)~\cite{talknet, TS-talknet, ava-as, activespeakersincontext}, where the camera and microphone are typically stationary relative to the scene rendering the recording conditions favourable. With the recent proliferation of wearable devices, however, the flavour of data \ac{ASD} systems are likely to be deployed upon has shifted to egocentric recordings, where the audiovisual signal is acquired from the first person perspective, and the camera and microphone are dynamic relative to the scene. This change in recording perspective introduces several challenges: (i) low signal-to-noise ratios for speech signals, (ii) highly spontaneous conversations with overlapping speech, (iii) audiovisual distortion caused by the camera wearer's head movements, and (iv) situational obfuscation, where visual cues are occluded~\cite{clarke23-ASD-ASRU, Ego4D}. 
Since the paradigm observed in recent literature involves modelling the correspondence between audiovisual cues indicative of a candidate speaker talking~\cite{EASEE,activespeakersincontext,ASDNet,Liao_2023_CVPR} (like lip movement, cheek posture~\cite{asdtransformer}, and audible speech), this paper argues said approaches are not sufficiently robust to handle the aforementioned challenges associated with egocentric recordings. For example, when speech is present in the audio signal but the video signal is heavily corrupted, a typical system based on audiovisual correspondence will only be able to recognise the presence of speech, the system will not be able to attribute it to the candidate speaker~\cite{activespeakersincontext}. This is illustrated in~\autoref{fig:timeline}: degraded video frames of the candidate speaker (bottom panel) and active speech in the microphone channel (top panel) that is not spoken by the candidate speaker induces a false activity detection for a speaker-embedding-naive system~\cite{Liao_2023_CVPR} (blue line in the middle panel of ~\autoref{fig:timeline}). This problem is also demonstrated by the disparity in performance when evaluating \ac{ASD} systems on exocentric~\cite{ava-as} vs egocentric~\cite{clarke23-ASD-ASRU, LoCoNet, SPELL} benchmarks, where in the latter, challenging scenes are regularly prevalent. 


Recent work has attempted to mitigate the limitations of established \ac{ASD} systems~\cite{talknet} by injecting speaker-specific information. Specifically, TS-TalkNet~\cite{TS-talknet} uses a pre-trained speaker recognition model to extract speaker embeddings from reference speech based on the well-known ECAPA-TDNN architecture~\cite{ecapa-tdnn}. These speaker embeddings are then leveraged as an additional source of information. Drawing inspiration from this, this paper proposes the \acf{SCAN}, an auxiliary module that can be integrated with various end-to-end \ac{ASD} systems \cite{talknet, Liao_2023_CVPR}. Unlike TS-TalkNet~\cite{TS-talknet}, \ac{SCAN} extracts speaker-specific information from two distinct sources: reference speech, i.e.~previously diarised speech spoken by the candidate speaker, and the candidate audio signal. This technique has previously leveraged successfully in the domain of personal voice activity detection~\cite{ding2022personalvad20optimizing, 10446042}. The novelty of \ac{SCAN} lies in its ability to perform framewise comparisons between these two sources via a cross-attention mechanism. This enables \ac{SCAN} to identify similarities and distinctions between speaker-specific cues in the reference speech and the candidate audio signal with high temporal granularity. By doing so, \ac{SCAN} provides a mechanism for effectively disambiguating scenarios with low-quality video signals, resulting in improved \ac{ASD} reliability and robustness. \ac{SCAN} constitutes the first contribution of this paper. Additionally, this work demonstrates that existing methods for generating identity-speech libraries, which associate reference speech with the identity of the candidate speaker, are not robust to the challenges of egocentric video. To address this, this paper proposes an improved method for generating identity-speech libraries by extending and finetuning an existing face-recognition model to leverage the temporal context of video data via a self-supervised learning objective. This constitutes the second contribution of this paper.



To summarise, the work outlined in this paper expands upon the concept of target-speaker \ac{ASD}, tailoring it to address the particular challenges posed by egocentric recordings with the following main contributions:

\newenvironment{tight_enumerate}{
\begin{enumerate}
  \setlength{\itemsep}{0pt}
  \setlength{\parskip}{0pt}
  \setlength{\topsep}{0pt}
}{\end{enumerate}}
\begin{tight_enumerate}
\item The auxiliary module \ac{SCAN}, which leverages speaker-specific information to help disambiguate challenging scenes for \ac{ASD}.
\item A self-supervised method to finetune a pre-trained face recognition model on video data to enroll identity-speech libraries more robust to egocentric recordings.
\item Performance analysis on the egocentric Ego4D-AVD dataset~\cite{Ego4D} and exocentric AVA-ActiveSpeaker~\cite{ava-as} using three existing systems as baselines: two speaker-embedding-naive systems - TalkNet~\cite{talknet} and Light-ASD~\cite{Liao_2023_CVPR}; and the current state-of-the-art speaker-embedding-informed system TS-TalkNet~\cite{TS-talknet}.
\end{tight_enumerate}

\section{Speaker Embedding Informed Audiovisual Active Speaker Detection}
\label{sec:method}

This section presents an overview of a typical \ac{ASD} architecture, introduces the proposed \ac{SCAN} module, and outlines the training protocol for fine-tuning an existing framewise face recognition model on egocentric video to construct more robust identity-speech (face-speaker\cite{TS-talknet}) libraries.

\begin{figure}[!ht]
  \centering
    \includegraphics[scale=0.05]{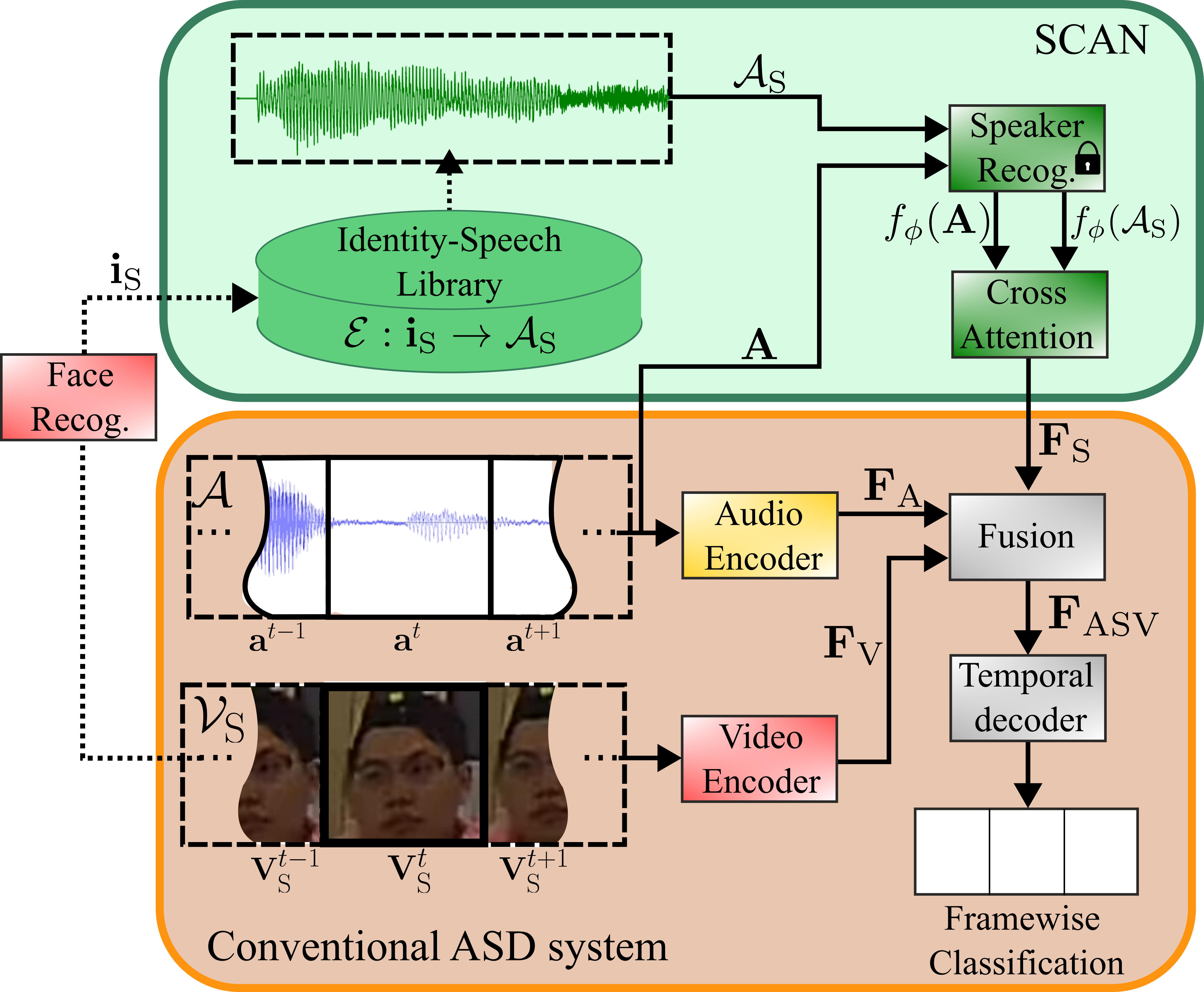}
  \caption{\ac{SCAN} is shown in the top box which leverages speaker-specific information for framewise comparison of reference speech and input audio signal via cross-attention. The bottom box shows a typical \ac{ASD} architectural design (baseline). Dotted connections represent non-end-to-end passages in the framework.}
\label{fig:ASD}
\end{figure}
\label{ssec:met_SREASD}


\vspace{-10pt}
\subsection{Baseline Architecture Overview}
As shown in the lower part of~\autoref{fig:ASD}, conventional \ac{ASD} systems, such as the speaker-embedding-naive systems used as baselines in this paper (i.e.~TalkNet~\cite{talknet} and Light-ASD~\cite{Liao_2023_CVPR}), follow the current paradigm of comprising an audio encoder, a video encoder, a modality fusion mechanism, and a temporal decoder~\cite{talknet, TS-talknet, Liao_2023_CVPR, ASDNet}. These systems operate on the video signal $\mathcal{V}_{\mathrm{S}}$ and the audio signal $\mathcal{A}$. The video signal is a set  $\mathcal{V}_\mathrm{S} = \{\mathbf{V}_{\mathrm{S},1}, ..., \mathbf{V}_{\mathrm{S}, T}\}$ of temporally contiguous video frames, centred on a single candidate speaker $\mathrm{S}$. Each grayscale image in the set is denoted $\mathbf{V}_{\mathrm{S},t} \in \mathbb{R}^{H\times W}$ with time index $t \in \{1, ..., T\}$, height dimension $H$, and width dimension $W$. $\mathcal{A} = \{\mathbf{a}_{1}, ..., \mathbf{a}_{T_\mathrm{A}}\}$ denotes the mixed audio signal temporally correspondent to $\mathcal{V}_{\mathrm{S}}$ with time index $t_{\mathrm{A}} \in \{1, ..., T_{\mathrm{A}}\}$. The distinction between $T$ and $T_{\mathrm{A}}$ compensates for the discrepancy in modality sampling rates.

Video and audio encoders extract pertinent features from their respective modality inputs, and perform short-term temporal modelling to encapsulate the local inter-frame relationships. These encoders typically resemble 3D-ResNets\cite{3DResNet}, \acp{V-TCN}\cite{talknet}, or depth-wise separable convolutions\cite{Liao_2023_CVPR}. The audio and video encoders embed their respective inputs, resulting in two matrices $\mathbf{F}_{\mathrm{A}} \in \mathbb{R}^{T \times d}$ and $\mathbf{F}_{\mathrm{V}}\in \mathbb{R}^{T \times d}$, where $d$ is the embedding dimension of both encoders. A fusion mechanism is then applied to combine these two embeddings, generating a single $2$-dimensional output $\mathbf{F}_{\mathrm{ASV}}$. This fusion is often a simple channel-wise concatenation, summation~\cite{Liao_2023_CVPR}, or an attention-based approach~\cite{talknet, LoCoNet}. The temporal decoder performs two tasks: long-term temporal modelling on the mixed-modality embedding to capture the sequential nature of speech and framewise classification to predict the speaker activity of the candidate speaker.


\subsection{Speaker Comparison Auxiliary Network (SCAN)}
The TS-TalkNet system, introduced in~\cite{TS-talknet}, builds upon the paradigm outlined in the bottom part of~\autoref{fig:ASD}. It does so by injecting speaker-specific information from pre-diarised reference speech of the candidate speaker $\mathcal{A}_{\mathrm{S}}$ into the model. A speaker embedding $f_{\phi}(\mathcal{A}_{\mathrm{S}})$ is extracted from $\mathcal{A}_{\mathrm{S}}$ via ECAPA-TDNN~\cite{ecapa-tdnn} which is then fused with $\mathbf{F}_{\mathrm{A}}$ and $\mathbf{F}_{\mathrm{V}}$ prior to temporal decoding. This injection of speaker-specific information relating to the candidate speaker results in a significant performance improvement over its respective baseline\cite{talknet}. This extension is motivated by the need to resolve ambiguous scenarios where visual cues indicative of speech activity are occluded, rendering the scenario intractable via audiovisual correspondence alone.

Inspired by this approach, \ac{SCAN} employs a distinct modification by extracting speaker embeddings from not only pre-diarised reference speech $\mathcal{A}_{\mathrm{S}}$, but also from the candidate audio signal $\mathcal{A}$. This enables an explicit comparison between speaker characteristics of the reference speech and the candidate audio signal via the cross-attention mechanism shown in~\autoref{fig:ASD}. As a result, it will be easier for the network to learn to identify similarities and distinctions between $\mathcal{A}$ and $\mathcal{A}_{\mathrm{S}}$.

First, overlapping windows temporally centred around each video frame are extracted from the raw waveform input of $\mathcal{A}$, transforming $\mathcal{A}$ to a matrix $\mathbf{A}$. 
This transformation is performed 
such that a sufficient duration of audio exists at each time point for meaningful speaker embeddings to be extracted. $\mathcal{A}_{\mathrm{S}}$ and $\mathbf{A}$ are then embedded by a pre-trained speaker recognition model $\mathrm{f}_\phi$ and a cross-attention mechanism is employed along the temporal dimension of $\mathbf{A}$: 

\begin{equation}
    \label{eq:crossatt}
    \mathbf{F}_{\mathrm{S}} = \sigma\left(\frac{\mathrm{f}_\phi(\mathbf{A})\mathrm{f}_\phi(\mathcal{A}_\mathrm{S})^\top}{\sqrt{d_\phi}}\right)\mathrm{f}_\phi(\mathcal{A}_\mathrm{S}) 
\end{equation} 
The embedded audio signal $\mathrm{f}_\phi(\mathbf{A})$ is used as the queries, and the embedded reference speech $\mathrm{f}_\phi(\mathcal{A}_\mathrm{S})$ is used as the keys and values. $d_\phi$ denotes the embedding dimension of the speaker recognition model and $\sigma$ represents the softmax function. This is done to determine how well the speech in the current track's audio correlates with that of the reference speech. ECAPA-TDNN~\cite{ecapa-tdnn} pre-trained on the VoxCeleb dataset~\cite{voxceleb} is used as $\mathrm{f}_\phi$ due to its previous use in TS-TalkNet and robust performance on various benchmarks~\cite{TS-talknet}. The ECAPA-TDNN model parameters are frozen and therefore do not contribute to model training. 

\vspace{-5pt}
\subsection{Identity-Speech Library Generation}
\label{sec:Identity-SpeechLibraryGeneration}

To exploit reference speech information for \ac{ASD}, a correspondence between candidate speaker identity and reference speech must be established. Following TS-TalkNet~\cite{TS-talknet}, this work generates an identity-speech library $\mathcal{E}: \mathbf{i}_{\mathrm{S}} \rightarrow \mathcal{A}_{\mathrm{S}}$ which is pre-enrolled offline, where $\mathbf{i}_{\mathrm{S}}$ is a vector representing the candidate speaker's identity. In \ac{ASD} datasets, identity annotations are typically not provided. However, by definition, tracks are identity-homogeneous. By clustering the identities of each track, this indirectly clusters each track's corresponding pre-diarised speech signal (if the track contains active speech). This is the premise of identity-speech library generation.

\subsubsection{Identity Aggregation}
\label{method:Identity_aggregation}
To construct the identity-speech library, identity embeddings $\mathbf{i}$ are extracted from the visual component of all tracks in a given dataset. Cosine similarity is used to assess the similarity between a pair of identity embeddings. If the similarity exceeds a static threshold, the identities within each track are considered to be the same, and pre-diarised speech within the track's corresponding audio signal is attributed to this identity in the library. 

\subsubsection{Self-Supervised Video-Based Face Recognition}
\label{ssec:VBFR}
The performance of the face recognition model used is critical to the quality of the identity-speech library and consequently the utility of the output of \ac{SCAN}. For example, if $\mathcal{E}:\mathbf{i}_{\mathrm{S}}$ contains speech not spoken by $\mathrm{S}$ when $\mathcal{A}_{\mathrm{S}}$ is sampled, irrelevant speech could be fed into the speaker recognition module, rendering its output uninformative, or even deleterious. The reliability of existing frame-based face recognition systems~\cite{Zheng2018RingLC, arcface, Liu2017SphereFaceDH, Wang2018CosFaceLM}, despite their robust performance on other benchmarks, will not be sufficient to withstand the challenges posed by the highly domain-specific nature of egocentric recordings.
Subsequently, a method was devised to adapt and finetune an existing pre-trained face recognition model. Firstly, a means of modelling consecutive frames as a sequence was integrated into a frame-based face recognition model, enabling it to effectively encapsulate and leverage the temporal context associated with video data. Secondly, since hard labels describing the trackwise identities of each person within \ac{ASD} datasets are not typically provided, a self-supervised training objective was employed. 



\begin{figure}[!ht]
  \centering
    \includegraphics[scale=0.05]{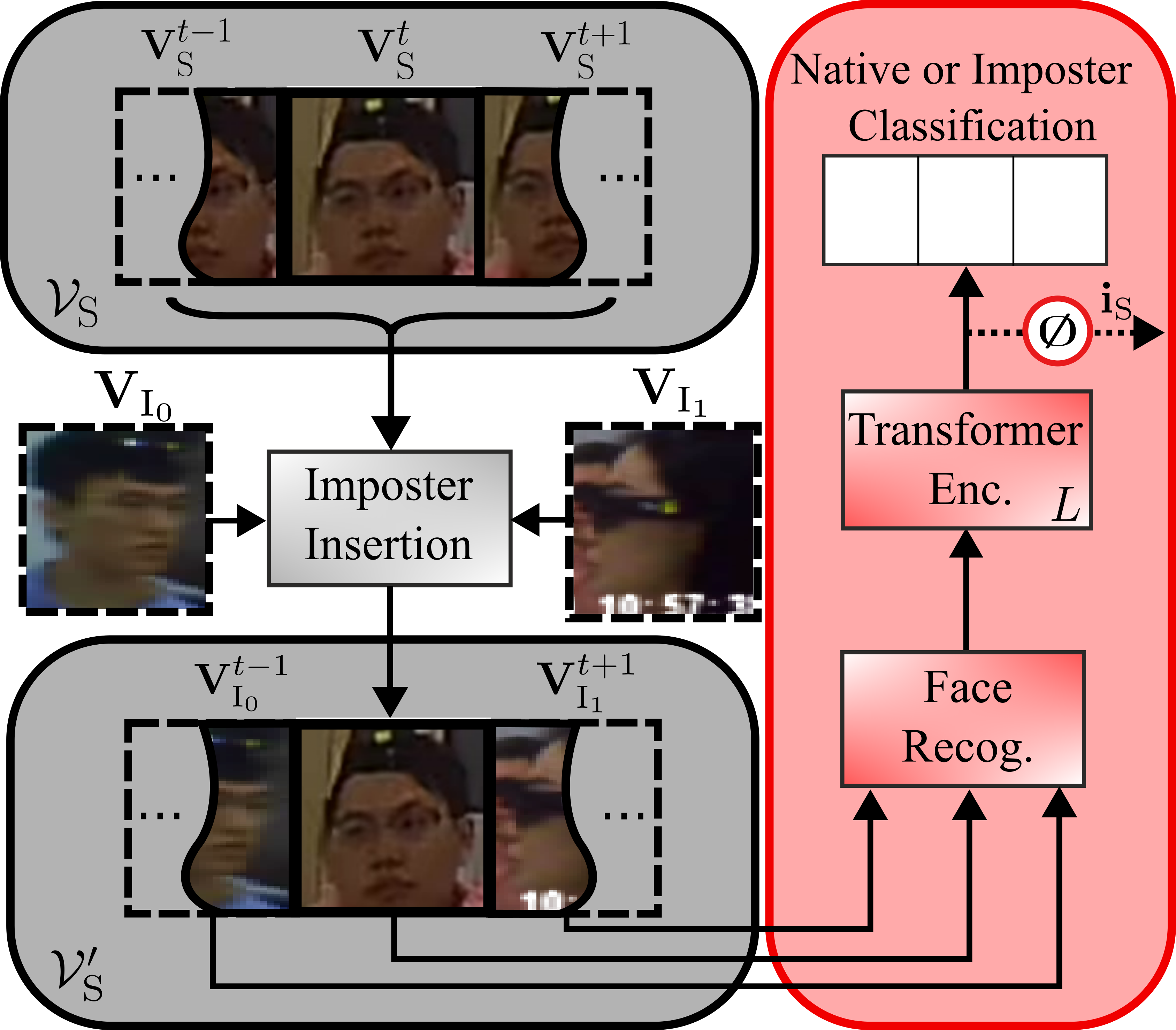}
  \caption{Self-supervised video-based face recognition model. impostor frames are randomly inserted into the parent track, resulting in polluted track $\mathcal{V}_{\mathrm{S}}'$. The training objective involves the model classifying frames as either native or impostor frames with respect to the parent track. $\diameter$ denotes mean average.}
\label{fig:VBFR}
\end{figure}

As depicted in~\autoref{fig:VBFR}, input tracks $\mathcal{V}_{\mathrm{S}}$ are polluted with impostor frames $\mathbf{V}_{\mathrm{I}_x}$ randomly chosen from other tracks. Frames within the polluted track $\mathcal{V}_{\mathrm{S}}'$ are then individually encoded by the face recognition model~\cite{glint360k}. This output is then fed through $L$ transformer encoder layers~\cite{attention} where the model has the capacity to attend across different frames within the track, thus leveraging the temporal context of video data. 
The model then determines whether each frame is either an impostor or native frame via binary classification. Learning this classification indirectly conditions the model to 
assign low weighting to frames not relevant to the track's overall identity (poor quality frames with significant visual distortion or occlusion) and high weighting to crisp frames where the parent track identity is clearly apparent and recognisable.

Once fine-tuned, the output of the last transformer encoder layer is averaged across its temporal dimension to generate a single embedding representative of the candidate speaker's identity $\mathbf{i}_{\mathrm{S}}$. 

\section{Experiments}
\label{sec:experiments}
This section describes the datasets and experiments used to evaluate \ac{SCAN} for \ac{ASD} and the quality of the identity-speech library.

\vspace{-5pt}
\subsection{Datasets}
\label{ssec:datasets}

\textbf{AVA-ActiveSpeaker}~\cite{ava-as} is a frequently-used, exocetric, large-scale audiovisual \ac{ASD} dataset, comprising $262$ Hollywood movie clips ($120$ for training, $33$ for validation, and $109$ for testing) with $3.65$ million human-labeled video frames ($38.5$ hours of face tracks) and corresponding audio.

\textbf{Ego4D-AVD}~\cite{Ego4D} records from the egocentric perspective. It comprises $572$ distinct video clips. Each video clip is $5$ minutes in length, some of which are recorded concurrently. All data is recorded monaurally using a variety of wearable devices. All video is sampled at $30$~Hz and uses high-definition resolution. The dataset is stratified as follows: $379$ clips for training, $50$ clips for validation, and $133$ clips for testing. The full validation fold of Ego4D-AVD was annotated by this work in terms of pseudo-identity. This was to provide a robust means of evaluating identity-speech libraries (cf. ~\autoref{tab:identity-speech}).

\subsection{Implementation Details}
\label{ssec:imps}

\textbf{Baselines:} All baseline models were implemented using the same input features, optimisers, and learning rates used in each systems's original implementations~\cite{talknet,Liao_2023_CVPR,TS-talknet}. Standard \ac{ASD} augmentation techniques were applied such as negative sampling of the audio signal, and flipping, cropping, and rotating of the video signal.

\textbf{SCAN:} The output of \ac{SCAN} contributed to each baseline system's loss function as an auxiliary loss, using binary cross entropy to perform framewise classification upon $\mathbf{F}_{\mathrm{S}}$. 
Raw waveform audio served as the reference speech input to the speaker recognition model ($1024$ channel ECAPA-TDNN~\cite{ecapa-tdnn}). The output of \ac{SCAN} $\mathbf{F}_{\mathrm{S}}$ used an embedding dimension of $64$. $1$ second windows of audio was used to extract each speaker embedding from $\mathbf{A}$. Reference speech was randomly selected from the relevant part of the identity-speech library to increase training variability. 

\textbf{Identity-Speech Library:} For the finetuning of the face recognition model (cf.~\autoref{ssec:VBFR}), a cross-entropy loss function was used. Input to the system were tracks comprising colour images with a $30$\% impostor insertion rate. $4$ Transformer encoder layers ($L=4$) each comprising $8$ attention heads with a model dimension of $1024$ were trained for $10$ epochs on a single NVIDIA A$100$ GPU for $2$ hours with a batch size of $1800$. To create the identity-speech library a static comparison threshold of $0.9$ was used to construct the library and a $2.5$ second minimum duration of speech was enforced.
\vspace{-5pt}
\subsection{Evaluation Metric}
\label{ssec:evaluation_metric}


Evaluation of each system for \ac{ASD} is performed using the Cartucho object detection \ac{mAP}\cite{cartucho}, which adheres to the \ac{mAP} criterion from the PASCAL VOC2012 competition\cite{pascal-voc-2012}. This approach is consistent with the Ego4D audiovisual diarisation challenge~\cite{Ego4D} and recent literature~\cite{clarke23-ASD-ASRU}. Due to the unavailability of ground truth annotations for the test folds of Ego4D and AVA-ActiveSpeaker, results are reported on the validation folds of each dataset, following the convention in \ac{ASD}~\cite{activespeakersincontext, ASDNet, clarke23-ASD-ASRU, SPELL, EASEE, LoCoNet}.
\vspace{-10pt}
\section{Results}
\label{sec:Results}
This section demonstrates the performance of \ac{SCAN} when used in conjunction with two speaker-embedding-naive systems, TalkNet~\cite{talknet} and Light-ASD~\cite{Liao_2023_CVPR}. The identity-speech library generation method proposed by this paper (cf.~\autoref{sec:Identity-SpeechLibraryGeneration}) is also evaluated and compared with previous work when applied to egocentric recordings.

\vspace{-7pt}

\subsection{Audiovisual Active Speaker Detection}
\label{ssec:results_ASD}
The results of incorporating \ac{SCAN} with two speaker-embedding-naive systems, TalkNet~\cite{talknet} and Light-ASD~\cite{Liao_2023_CVPR}, are shown in~\autoref{tab:speaker_embeddings}. 

\vspace{-5pt}
\begin{table}[!ht]
\centering
\caption{Performance comparison on Ego4D-AVD and AVA validation fold. Identity-Speech Library$\dagger$ refers to ground truth identity-speech library. Bold highlights best-performing system with hypothesised identity-speech library, underlined represents best system with ground truth identity-speech library.}
\label{tab:speaker_embeddings}
\begin{tabular}{ccccc}
\hline
{} & \textbf{SCAN} & \textbf{Identity-Speech } & \multicolumn{2}{c}{\textbf{mAP {[}\%{]}}} \\
\multirow{-2}{*}{\textbf{Baseline}} & \textbf{used} & \textbf{Library$\dagger$} & \textbf{Ego4D} & \textbf{AVA} \\ \hline
{}              &\ding{55} & \ding{55} & 52.2                        & \textbf{93.9}                      \\
\multirow{-2}{*}{TS-TalkNet}     & \ding{55}& \checkmark & 54.0                        &      93.9                     \\
\hline
{}                 &\ding{55} &- & 51.0                        & 92.3                      \\
{TalkNet}           & \checkmark & \ding{55} & \textbf{58.0}               &            93.8               \\
{}   & \checkmark & \checkmark & 58.4                        &                    94.0       \\ 
\hline
{}              &\ding{55} & - & 54.3                        & 94.1                      \\
Light-ASD          & \checkmark & \ding{55}& 57.1                        &                \textbf{93.9}           \\
{} & \checkmark & \checkmark & {\ul 59.9}                  &                {\ul 94.2}  \\ \hline
\end{tabular}
\end{table}

On the Ego4D benchmark, \ac{SCAN} significantly improves performance of the respective baseline systems for both ground truth identity-speech library and hypothesised identity-speech library configurations. Ground truth identity-speech libraries referring to those which are created directly from the dataset's annotation, hypothesis identity-speech library referring to those created by the method outlined in~\autoref{sec:Identity-SpeechLibraryGeneration}. For the AVA benchmark (exocentric) the improvements are much more modest. This is likely because visually challenging multi-talker scenarios, in which \ac{SCAN} would be beneficial, are much less prevalent than in Ego4D. Nevertheless, both configurations provide a substantial improvement upon the TalkNet baseline system. Additionally, TalkNet+\ac{SCAN} outperforms TS-TalkNet, a previous speaker-embedding-informed system by $5.8$\% and $4.4$\% \ac{mAP} for ground truth and non-ground truth identity-speech libraries, respectively. Since a significant improvement upon the TS-TalkNet baseline is apparent when ground truth identity-speech libraries are used, it is fair to deduce \ac{SCAN}'s architectural implementation and method of extracting speaker-specific information from both the candidate audio signal and reference speech is more effective than relying solely on reference speech. 
Furthermore, the improvement yielded by incorporating \ac{SCAN} into the baseline systems renders both baseline systems almost competitive with state-of-the-art methods in the context of egocentric data. Specifically, \ac{SCAN} enhances the TalkNet and Light-ASD baselines by $14.5\%$ and $10.3\%$, respectively, bridging the gap with state-of-the-art performance, as shown in~\autoref{tab:sota}

\vspace{-5pt}
\begin{table}[!ht]
\centering
\caption{Comparison with the state-of-the-art \ac{ASD} systems on validation folds of Ego4D and AVA. Values for LoCoNet~\cite{LoCoNet} and SPELL~\cite{spellego4dchallenge} are from their original manuscripts.}
\label{tab:sota}
\begin{tabular}{cccc}
\hline
\textbf{System}      & \textbf{Spk. Emb. Inf.} & \textbf{Ego4D {[}\%{]}} & \textbf{AVA {[}\%{]}} \\ \hline
TalkNet~\cite{talknet}              & \ding{55}                        & 51.0                        & 92.3                      \\
TS-TalkNet$\dagger$           &  \checkmark                        & 54.0                        & 93.9                      \\
Light-ASD~\cite{Liao_2023_CVPR}            &   \ding{55}                      & 54.3                        & 94.1                      \\
LoCoNet~\cite{LoCoNet}              &     \ding{55}                    & 59.7                        & \textbf{95.2}             \\

SPELLL~\cite{spellego4dchallenge}                &   \ding{55}                      & \textbf{60.7}               & 94.2                      \\ \hline
 TalkNet+SCAN$\dagger$   &  \checkmark                       & 58.4                        &    94.0                       \\
 Light-ASD+SCAN$\dagger$ &   \checkmark                      & 59.9                        &    93.9   \\ \hline
\end{tabular}
\end{table}
\vspace{-10pt}

\subsection{Identity-Speech Library}
\label{ssec:results_ISL}

\begin{figure}[!ht]
  \centering
    \includegraphics[scale=0.062]{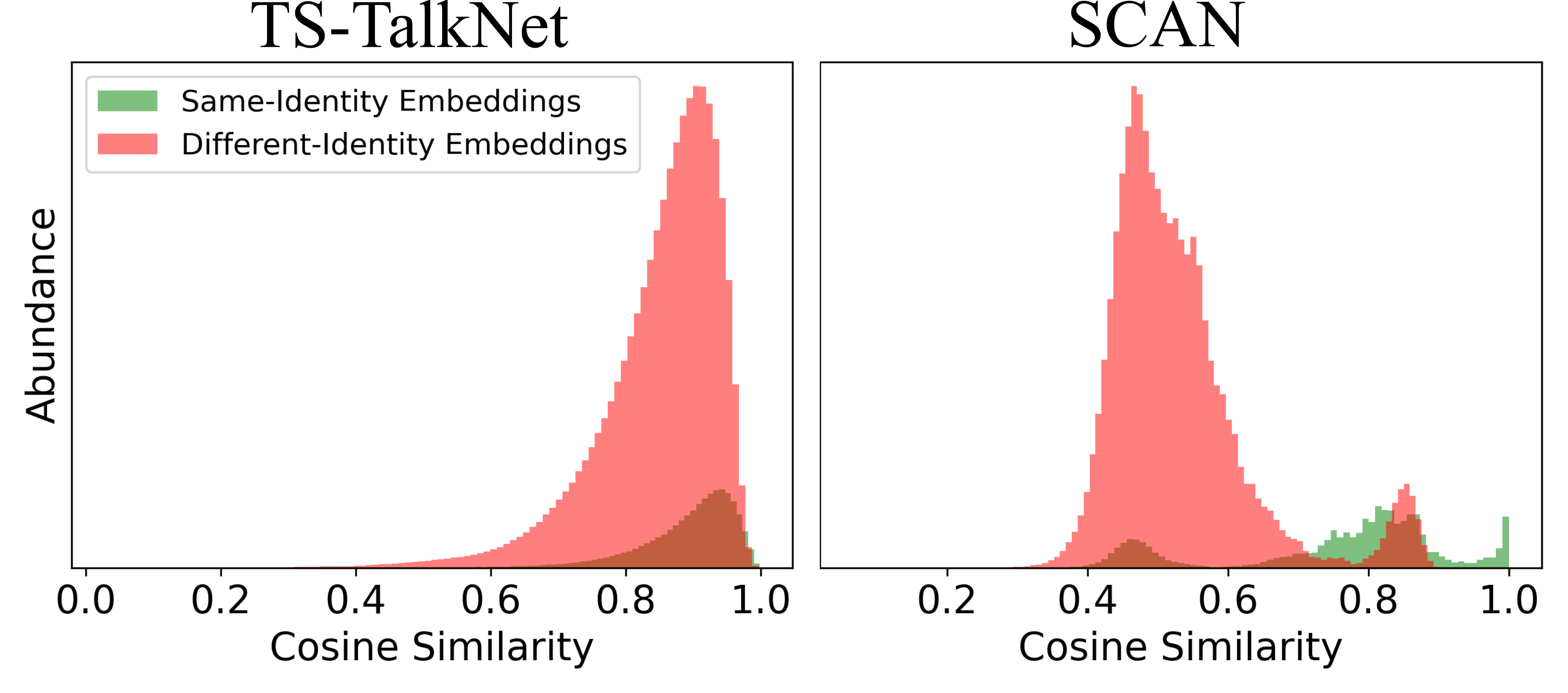}
  \caption{Similarity between same-identity embeddings and different-identity embeddings shown in green and red, respectively, for Ego4D validation fold~\cite{Ego4D}}
\label{fig:hist}
\end{figure}

The results presented in~\autoref{tab:identity-speech} indicate a substantial improvement in the quality of the identity-speech library generated by the proposed method. This improvement is further demonstrated by~\autoref{fig:hist}. In the left panel (TS-TalkNet), it is impossible to differentiate same-identity pairs from different identity pairings while in the right panel (\ac{SCAN}) resolving the two pairings is easier. This is attributed to the face-recognition model's ability to leverage temporal context via self-attention. However, it is noted that the silhouette score of $0.16$ indicates only minor cluster separability, suggesting that further refinement of the proposed method might be necessary to achieve more robust future identity-speech library generation.
\vspace{-5pt}
\begin{table}[!ht]
\centering
\caption{Comparison of identity-speech library generation methods.}
\label{tab:identity-speech}
\begin{tabular}{cc}
\hline
\textbf{System} & \textbf{Ego4D-Silhouette} \\ 
\hline
TS-TalkNet      & -0.17            \\
SCAN            & \textbf{0.16}     \\ \hline
\end{tabular}
\end{table}
\vspace{-10pt}
\vspace{-10pt}




\section{Conclusion}

This work proposes \ac{SCAN}, a speaker-embedding-informed extension to conventional \ac{ASD} systems. \ac{SCAN} assists in disambiguating challenging multi-talker scenarios involving visual noise and physical obfuscations. \ac{SCAN} builds upon previous work by extracting speaker-specific information from reference speech, but is able to leverage speaker-specific information inherently present in the candidate audio signal itself. Furthermore, \ac{SCAN} proposes a method to finetune frame-based face-recognition models on video data without hard identity labels by transformer encoder layers and a self-supervised training objective. This approach exhibits a significant performance improvement relative to previous work for identity-speech library generation.


\bibliographystyle{IEEEtran}
\bibliography{mybib}


\end{document}